\documentstyle[preprint,aps,epsf]{revtex}
\tighten

\begin{document}

\draft

\title{
Sensitivity Studies for Extraction of $\bbox{G_E^n}$ from Inclusive and
Semi-inclusive Electron Scattering on Polarized $\bbox{^3}$He.
}

\author{
J.~Golak$^{1,2}$,
W.~Gl\"ockle$^1$,
H.~Kamada$^3$
H.~Wita\l{}a$^2$, 
R.~Skibi\'nski$^2$,
A.~Nogga$^4$
}
\address{$^1$Institut f\"ur Theoretische Physik II,
         Ruhr Universit\"at Bochum, D-44780 Bochum, Germany}
\address{$^2$Institute of Physics, Jagiellonian University,
                    PL-30059 Cracow, Poland}
\address{$^3$ Department of Physics, Faculty of Engineering,
   Kyushu Institute of Technology,
   1-1 Sensuicho, Tobata, Kitakyushu 804-8550, Japan}
\address{$^4$ Department of Physics, University of Arizona, Tucson, 
              Arizona 85721, USA}
\date{\today}
\maketitle

\begin{abstract}
The processes 
$\overrightarrow{^3{\rm He}}(\vec{e},e')$ and
$\overrightarrow{^3{\rm He}}(\vec{e},e'n)$
are theoretically analyzed with
the aim to search for sensitivities in the electric form factor of the
neutron, $G_E^n$. Faddeev calculations based on the high precision NN
force AV18 and using consistent MEC's are employed. While the inclusive
process is too insensitive, the semi-exclusive one appears promising.
\end{abstract}

\narrowtext

\section{Introduction}
\label{secIN}

The experimental knowledge of electromagnetic form factors of the neutron
is of basic interest for testing model or finally QCD based predictions.
Quite intensive experimental efforts are planned~\cite{Alarcon} 
and have been undertaken to extract
these form factors from electron scattering on the 
deuteron~\cite{deJager,Anklin.98,Passchier.99,Herberg.99} 
and $^3$He~\cite{Becker.99,Rohe.99,Xu.00,Golak.01}.
In \cite{Xu.00} the magnetic neutron form factor $G_M^n$ has been extracted from
the process $\overrightarrow{^3{\rm He}}(\vec{e},e')$
at $q^2$= 0.1 and 0.2 ${\rm (GeV/c)}^2$.
The analysis of the
data relied on precise solutions of the 3N Faddeev equations for $^3$He and
the 3N continuum, thereby using modern nuclear forces and consistent
mesonic exchange currents. The resulting values for $G_M^n$ agreed
perfectly with results extracted from the cross section ratio
$ d(e,e'n) / d(e,e'p) $ \cite{Anklin.98}.
The experimental
data for higher $q^2$-values have not yet been analyzed in the same
framework because it has to be expected that relativistic corrections
will play a significant role and the theoretical framework for that
extension  has not yet
been settled enough to be reliably applicable. This is an important
challenge and task for theory.

In the case of $G_E^n$ the experiments~\cite{Becker.99,Rohe.99} for the process
$\overrightarrow{^3{\rm He}}(\vec{e},e'n)$
had the aim to extract
the electric form factor of the neutron.
The analysis, however, leaves more questions of reliability 
open than in case
of $G_M^n$. Around
$q^2$= 0.35 ${\rm (GeV/c)}^2$
a first result \cite{Becker.99} was based on the simple
assumption that polarized $^3$He can be considered to be a polarized neutron. 
This
was later corrected by a Faddeev calculation \cite{Golak.01}, 
however without taking
MEC's into account. Also relativistic effects in that Faddeev
calculation were not included, though they might be not negligible.
The corrections induced by final state interactions (FSI)
turned out to be substantial and moved the original
value towards the region of $G_E^n$-values found in the experiments
based on a deuteron target~\cite{Passchier.99,Herberg.99}. 
The theoretical analysis of that experiment~\cite{Golak.01}
was also aggravated by a heavy averaging over the experimental
conditions. 
At an even higher $q^2$-value of $q^2$= 0.67 ${\rm (GeV/c)}^2$
the same process was again
used under the same assumption of replacing $\overrightarrow{^3{\rm He}}$
by a polarized neutron
 to extract a value
of $G_E^n$~\cite{Rohe.99}. 
Corrections coming from a full $^3$He wave function
and rescattering
processes have not yet been estimated.

In such a situation it is of interest to theoretically investigate
electron induced $^3$He observables with respect to their sensitivity to
$G_E^n$. The ideas \cite{Blankleider} for choosing certain observables 
are based on  plane wave impulse approximation and the fact 
that the polarization of $^3$He is
carried with about 90 \% by the polarized neutron.
Thus it  is well known (see for instance~\cite{Ishikawa.98}) that 
under neglection of FSI and keeping only the principal S-state
an
asymmetry based on scattering of  a polarized electron on a $^3$He target
polarized perpendicular to the (virtual) photon direction is
proportional to $G_E^n G_M^n$. 
In~\cite{Schulze} inclusive scattering has been investigated 
under the assumption of PWIA but keeping a full $^3$He wave function 
with the pessimistic result, 
that the proton contribution overwhelms the signature of $G_E^n$.
The question remains what happens under the full dynamics.
Based on the same simple picture
one can form a ratio of two asymmetries, one with the $^3$He spin
perpendicular and one parallel to the photon direction. That ratio will
be proportional to $G_E^n/G_M^n$. In order to focus more on the neutron
one uses the $\overrightarrow{^3{\rm He}}(\vec{e},e'n)$
reaction and measures the knocked out neutron in
coincidence with  the scattered electron. Again the important question
arises: will sensitivity to $G_E^n$ remain when the full dynamics 
is taken into account ?

We investigate these questions using full fledged Faddeev calculations
and modern nuclear forces and including MEC's as well. 
We restrict ourselves to a strictly
nonrelativistic treatment even if we go into higher $q^2$-ranges, where
relativity should and will play a role. At least we can 
get insight into the importance or decrease of importance of FSI.

The paper is organized as follows. In Sec. II we briefly review the
theoretical framework. Our results for inclusive scattering 
and for the semi-exclusive processes are shown in Sec. III.
We summarize in Sec. IV and end with an outlook.

\section{Theoretical framework}
\label{secII}

The cross section for the process  
$\overrightarrow{^3{\rm He}}(\vec{e},e')$ is given as
 ~\cite{Donnelly}

\begin{equation}
{ {d^{\, 3} \sigma} \over { d \hat{k}' \, d k_0' } } \ = \
\sigma_{\rm Mott} \,
\left\{ v_L R^L + v_T R^T
\, + \, h \left( v_{TL'} R^{TL'} + v_{T'} R^{T'} \right) \right\} ,
\label{eq:sigma3}
\end{equation}
where $k_0'$, $\hat{k}'$ are the energy and direction of the scattered electron,
$v_L$, $v_T$, $v_{TL'}$, $v_{T'}$ 
are kinematical factors, 
$R^L$, $R^T$, $R^{TL'}$, $R^{T'}$ response functions and $h$ the
helicity of the initial electron. The asymmetry is defined as

\begin{equation}
A \ = \
{
{
\left. { {d^{\, 3} \sigma} \over { d \hat{k}' \, d k_0' } }\right|_{h=1}
\ - \ \left. { {d^{\, 3} \sigma} \over { d \hat{k}' \, d k_0' } }\right|_{h=-1}
} \over
{
\left. { {d^{\, 3} \sigma} \over { d \hat{k}' \, d k_0' } }\right|_{h=1}
\ + \ \left. { {d^{\, 3} \sigma} \over { d \hat{k}' \, d k_0' } }\right|_{h=-1}
}
}  \ = \ -
{
{
 v_{T'} \tilde{R}^{T'} \cos \theta^\star
\, + \, 2 \, v_{TL'} \tilde{R}^{TL'} \sin \theta^\star \cos \phi^\star 
} \over
{
 v_{L} R^{L} \, + \, v_{T} R^{T} 
}
} , 
\label{eq:ANav}
\end{equation}
where the dependence on $\theta^\star$ and $\phi^\star$ has been shown
explicitly. These angles denote the direction of the $^3$He spin
in relation to the direction of the photon. 
(In contrast to \cite{Ishikawa.98} we modified slightly the definition 
of the $\tilde{R}$ responses.) 
In \cite{Ishikawa.98} 
it is shown that in PWIA and under the
assumption of keeping only the principal $S$-state of the $^3$He wave
function that asymmetry is given as

\begin{eqnarray}
A^{\rm PWIA} \ = \ &
\frac{
-\frac{q^2}{2 m_N^2} \, \tan {\Theta \over 2} \left[
\sqrt{{q^{2} \over {\vec Q ^{2}} } + \tan^{2} {\Theta \over 2}}
( G_M^{(n)} )^2 \, \cos \theta ^{*} \ + \
\frac{2 m_N}{\vert \vec Q \vert} \,
F_1^{(n)}
G_M^{(n)} \,
\cos \phi ^{*} \sin \theta ^{*} \right]
}
{
( F_1^{(n)} )^2
 + 2 ( F_1^{(p)} )^2
 - \frac{q^2}{4 m_N^2}
\left[
( G_M^{(n)} )^2 + 2 ( G_M^{(p)} )^2 +
{\alpha}
\frac{6 m_N^2}{{\vert \vec Q \vert}^2 }
\left( ( F_1^{(n)} )^2 + 2 ( F_1^{(p)} )^2 \right)
\right] ( 1 + 2 \tan^{2} {\Theta \over 2} )
} .
\label{eq77} 
\end{eqnarray}
There is a reminder of the $^3$He wave function, the quantity $\alpha$, 
which, however, is numerically insignificant~\cite{Ishikawa.98}.
We now replace $F_1^n$ in the charge density operator by $G_E^n$.
Because of the smallness of $F_1^n$ the "relativistic correction" 

\begin{eqnarray}
G_E^n = F_1^n - { q^2 \over {4 m_N^2} } (G_M^n - F_1^n) \approx 
F_1^n - { \mid \vec{Q} \mid ^2 \over {4 m_N^2} } (G_M^n - F_1^n)
\label{eqGEn} 
\end{eqnarray}
is mandatory. (Please note
a misprint in Eq.~(78) of \cite{Ishikawa.98}: the square bracket 
in the denominator
should  end not before but behind 
$\tan^{2} {\Theta \over 2} $).
Our notation for the photon momentum is 
$ Q = ( \omega, \vec Q ) $ and $ -Q^2= q^2 = {\vec Q}^{\, 2} -\omega^2 $.
Regarding Eq.~(\ref{eq77}) we see
that $\theta ^{*}$= 0 $^\circ$ (90 $^\circ$) emphasizes $(G_M^n)^2$ ($G_E^n G_M^n$). 
In the present 
investigation we shall study the dependence of 
$A_\perp \equiv  A ( \theta ^{*}= 90 ^\circ , \phi^{*}= 0 ^\circ ) $
on $G_E^n$ including FSI
and MEC's. We shall also provide insight into the
contributions arising from photon absorption on the protons. 
This extends first studies
carried through in \cite{Ishikawa.98}, where only FSI effects 
were investigated.

The second process we are going to study is 
$\overrightarrow{^3{\rm He}}(\vec{e},e'n)$.
The sixfold differential cross section 
is given as ~\cite{Donnelly}

\[
{ {d^{\, 6} \sigma} \over { d \hat{k}' \, d k_0' \, d \hat{p}_n d E_n} } \ = \
{\cal C} \, \sigma_{\rm Mott} \, p_n \, { { p \, m_N^2} \over 2 }
\]
\begin{eqnarray}
\times \int d \hat{p}
\left\{ v_L R^L + v_T R^T + v_{TT} R^{TT} + v_{TL} R^{TL}
\, + \, h \left( v_{TL'} R^{TL'} + v_{T'} R^{T'} \right) \right\} ,
\label{eq:sigma6}
\end{eqnarray}
where in addition to what has been said before $\hat{p}_n$, $p_n$, $E_n$, 
$p$, $\hat{p}$ denote
the neutron direction, its momentum, its (nonrelativistic) kinetic energy,
the magnitude of the relative momentum of the two undetected protons 
and its direction. ${\cal C}$= $\frac12$ for two undetected protons.
Note that ${\cal C}$= 1 for the $\overrightarrow{^3{\rm He}}(\vec{e},e'p)$
reaction. 

Now the asymmetry defined in the  same manner in relation to the
electron helicities is given as
\begin{eqnarray}
A \ = \
{ { \int d \hat{p} \left(  v_{T'} R^{T'} + v_{TL'} R^{TL'} \right) } \over
{ \int d \hat{p} \left( v_L R^L + v_T R^T + v_{TT} R^{TT} + v_{TL} R^{TL} \right) } }
\label{eq:Asymmetry}
\end{eqnarray}

We  form the ratio  $ A_\perp / A_\parallel $,
where $ A_\perp (A_\parallel) $ refers to 
$ \theta ^{*}$= 90 $^\circ$ (0 $^\circ$) and study its sensitivity
to changes in $G_E^n$ and FSI
as well as  MEC influences. It will also be of interest to see the proton
contribution to that ratio, which is mostly caused by rescattering.

The technical performance in momentum space and the necessarily
involved partial wave decomposition has been described 
in~\cite{Golak.01} and references therein.

\section{Results}
\label{secIII}

\begin{figure}[h!]
\leftline{\mbox{\epsfxsize=120mm \epsffile{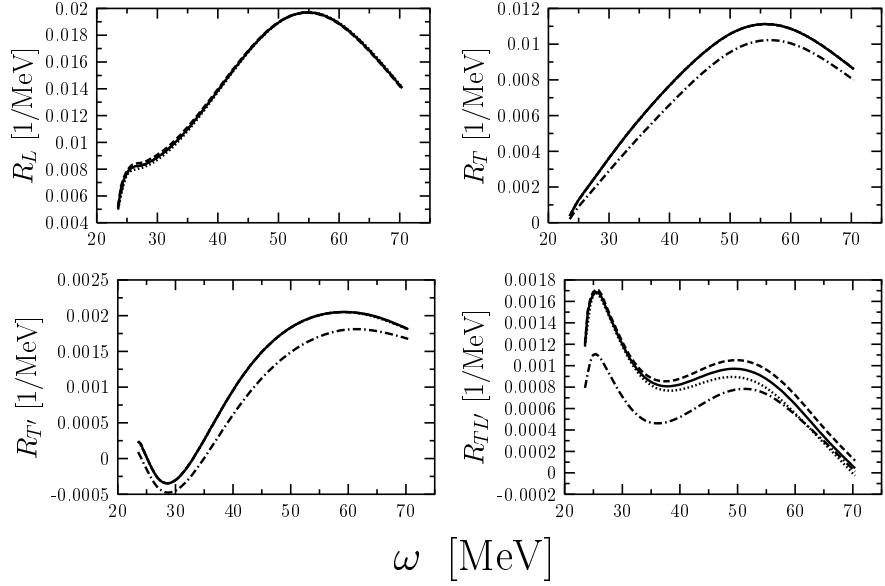}}}
\caption[ ]
{$R_L$, $R_T$, $R_{T'}$ and $R_{TL'}$
for $q^2$= 0.1 ${\rm (GeV/c)}^2$. 
``Full'' (solid), FSI without MEC (dash-dotted), ``Full'' with 1.6$G_E^n$ (dashed)
and with 0.4$G_E^n$ (dotted).
}
\label{fig1}
\end{figure}

\begin{figure}[h!]
\leftline{\mbox{\epsfxsize=120mm \epsffile{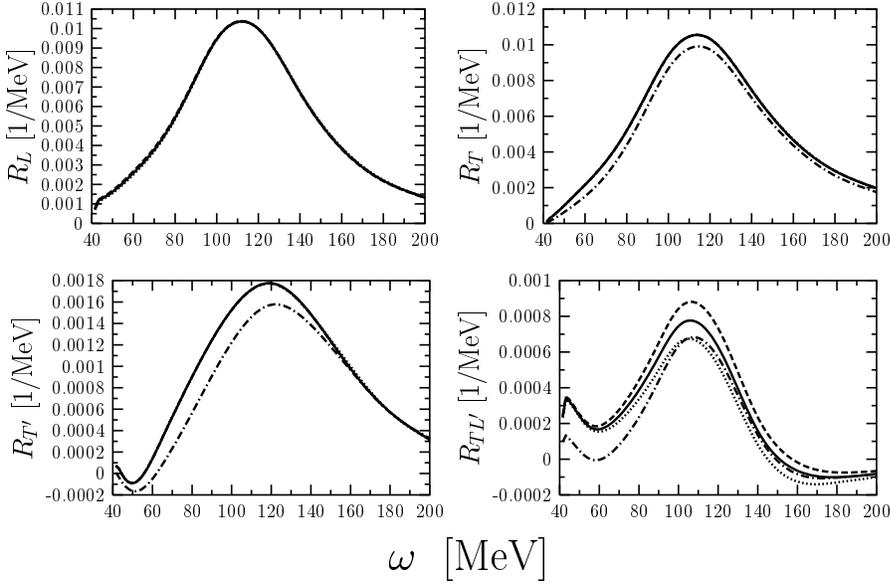}}}
\caption[ ]
{The same as in Fig.~\protect\ref{fig1}
for $q^2$= 0.2 ${\rm (GeV/c)}^2$.
}
\label{fig2}
\end{figure}

We first regard the process $\overrightarrow{^3{\rm He}}(\vec{e},e')$. 
Throughout we use the high
precision NN force AV18~\cite{AV18} together with $\pi$- and $\rho$-like 
MEC's~\cite{Kotlyar} according to
the Riska prescription~\cite{Riska}. 
As a
reference model we take the H\"ohler parametrisation for all electromagnetic
form factors of the nucleons~\cite{Hoehler}. 
There are more recent parametrizations, which are based on newer data,
fulfill constraints of pQCD etc.~\cite{Meissner}, which, however,
would not change the conclusions of our study.
Besides the neglection of relativistic corrections the insufficient
knowledge of the MEC's appears to be a second concern about theoretical
uncertainties. While the NN force  chosen has been at least adjusted to the
rich set of NN data the choice of MEC's is  not  constrained in a
corresponding manner. The ones we are using are however at least in
harmony with the continuity equation. In view of this situation we
would like to show results without and with inclusion of MEC's. Thus one
can see the magnitudes of the shifts caused by the MEC's alone. 
The calculations including FSI and MEC's will be denoted by ``Full'' 
in the following. What we call PWIAS does not include FSI nor MEC's but 
allows photon absorption on all three nucleons. 
This can also be expressed as photon absorption, say on nucleon~1, 
but then keeping fully antisymmetrized plane waves in the final state.
We show in Figs.~\ref{fig1} and \ref{fig2}  
the four response functions $R_L$, $R_T$, $R_{T'}$ and $R_{TL'}$ 
as a function of the energy transfer $\omega$. 
The first (second) case shown in  Fig.~\ref{fig1} (Fig.~\ref{fig2}) 
corresponds roughly to $q^2$= 0.1 (0.2) ${\rm (GeV/c)}^2$. 
More precisely in the two cases we have chosen the initial 
electron energy to be 
778 (1728) MeV and the electron scattering angle as 23.7 (15.0) $^\circ$.
There are always four curves: one is the reference curve with the
$G_E^n$ as given in~\cite{Hoehler} and ``Full'' dynamics 
and another one with FSI but without MEC's. 
The two other curves are of ``Full'' type but $G_E^n$ is multiplied
by 1.6 and 0.4, respectively.
$R_L$ is not 
affected by MEC's since we do not
include two-body densities. 
Its dependence on $G_E^n$ is marginal, since $R_L$ is dominated by the
proton.
Besides into the density operator $G_E^n$ also
enters into the MEC's, but there only as a difference to the proton form
factor. Consequently changes of $G_E^n$ hardly affect $R_T$ and $R_{T'}$.
Still the both response functions are visibly changed by MEC's.
$R_{TL'}$ is the only response function of interest in searching 
for $G_E^n$ sensitivities. 
Unfortunately we see quite a strong effect of MEC's which introduces a
theoretical uncertainty as mentioned in the introduction. For the MEC's
chosen the $ \pm $ 60~\% changes in $G_E^n$ lead to about $ \pm $
8~\% changes in $R_{TL'}$ in its quasi-elastic peak region around 
$ \omega$ = 50 MeV. This is  for $q^2$= 0.1 ${\rm (GeV/c)}^2$.

For $q^2$= 0.2 ${\rm (GeV/c)}^2$ those changes are larger.
They amount to $\pm $  13 \% in the quasi-elastic peak region around 
$\omega$= 100 MeV. 
This is highly insufficient to serve as a signature for $G_E^n$. 
On top there are uncertainties coming from MEC's.
The reason
for these small changes lies in the strong proton contribution as already
shown in~\cite{Schulze}, based, however, on a PWIA calculation. This is now
confirmed using the full dynamics. 

We performed one calculation for each $q^2$ in the peak region dropping 
all proton electromagnetic form factors. 
The results are shown in Table I. We see that $R_L$ is totally
dominated by photon absorption on the proton. The reductions for $R_T$ by
switching off the proton contribution are about 87 \% , while they
are much less for $R_{T'}$, namely about 27 \% . Now in case of $R_{TL'}$
one has reductions of 87 \% and 80 \% at $q^2$= 0.1 and 0.2 ${\rm (GeV/c)}^2$,
which explains the insufficient sensitivity against changes in
$G_E^n$ at these $q^2$-values. We refrained from investigating  higher 
$q^2$-values because missing relativistic effects might change the results.  
There is no need to compare with PWIAS calculations,
since they are known~\cite{Golak.95} to be insufficient.

\begin{figure}[h!]
\leftline{\mbox{\epsfxsize=120mm \epsffile{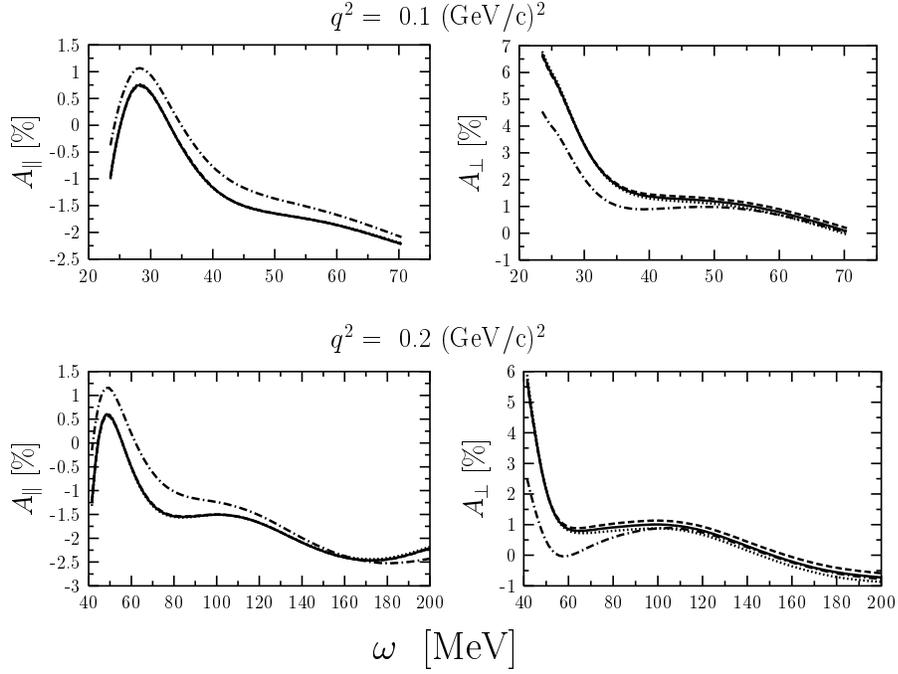}}}
\caption[ ]
{
$A_\parallel$ and $A_\perp$
for $q^2$= 0.1 and 0.2 ${\rm (GeV/c)}^2$.
Curves as in Fig.~\protect\ref{fig1}.
}
\label{fig3}
\end{figure}

Since for those changes of $G_E^n$ the shifts in $R_L$ and $R_T$ are
negligible 
the changes in the asymmetry $A_\perp$ reflect
directly the changes in $R_{TL'}$. This is shown in Fig.~\ref{fig3},
which for the sake of completeness also includes $A_\parallel$. 
We see first of
all the strong shifts caused by the MEC's. Then around $\omega$= 50 (100) MeV
for $q^2$= 0.1 (0.2) ${\rm (GeV/c)}^2$ 
small modifications of 
$A_\perp$ of about $\pm$  8 (13) \% are seen caused by the $\pm$ 60 \%
variations in $G_E^n$. The strong proton contribution  
explains the insufficient sensitivity against $G_E^n$.

\begin{figure}[h!]
\leftline{\mbox{\epsfxsize=120mm \epsffile{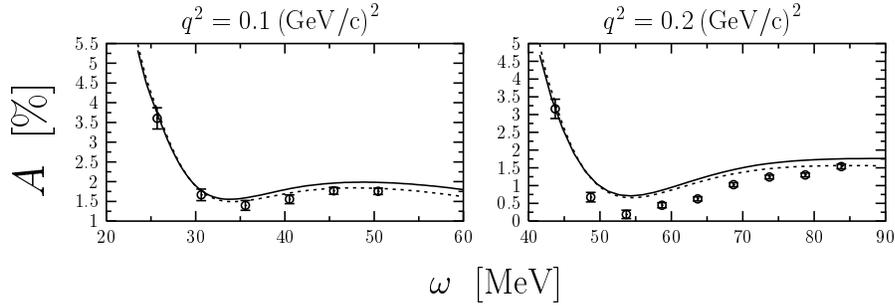}}}
\caption[ ]
{
Comparison of data~\cite{Xiong.01} at $q^2$= 0.1 and 0.2
${\rm (GeV/c)}^2$ with two ``Full'' point geometry calculations 
using $F_1^n$ (dashed) and $G_E^n$ (solid), respectively.
}
\label{fig4}
\end{figure}

The measurements of both asymmetries are nevertheless of great
importance. $A_\parallel$  has been used recently~\cite{Xu.00} 
to extract $G_M^n$ as mentioned
in the introduction. Pioneering measurements on the asymmetry $A_\perp$ 
have been performed in~\cite{aperp}.
They have been analyzed in~\cite{Ishikawa.98},  however
without MEC's and using $F_1^n$ instead of $G_E^n$ in the single nucleon
density operator. The agreement with those data was quite good. More
recently the asymmetry given in Eq.~(2) was measured around 
$\theta^\star$ = 130 to 140$^\circ$~\cite{Xiong.01}.
We analyzed the data with calculations of the ``Full'' type.
The agreement was quite good at $q^2$= 0.1  ${\rm (GeV/c)}^2$
but an overshooting of the
theory was observed for $q^2$= 0.2 ${\rm (GeV/c)}^2$.
It has to be remarked that
in those calculations still $F_1^n$ has been used in the single nucleon
density operator. Despite the fact that a strong proton contribution is
present the changes by going from $F_1^n$ to $G_E^n$ are noticeable. We
document that in Fig.~\ref{fig4} by comparing the data at $q^2$= 0.1 and 0.2
${\rm (GeV/c)}^2$~\cite{Xiong.01} 
with two ``Full'' calculations using $F_1^n$ and $G_E^n$, respectively.
Using $G_E^n$ leads to a slight deterioration in comparison to the $F_1^n$ 
result.

Let us now move on to the process $\overrightarrow{^3{\rm He}}(\vec{e},e'n)$
and check whether it is more sensitive to $G_E^n$.
As emphasized before our
present strictly nonrelativistic framework does not allow reliable
predictions at high $q^2$-values, say above $q^2$= 0.2 ${\rm (GeV/c)}^2$.
Nevertheless we shall now exhibit results beyond that
with the only aim to describe possible trends for the significance of FSI and
MEC's. We cannot exclude that these results might change  in the future 
to an unknown extend, when relativity will be correctly included.

With respect to extracting neutron information it appears optimal to choose
a break up configuration where the neutron is knocked out in the
direction of the photon. On top one can assume that the neutron receives
the full photon momentum and moreover the photon energy equals the final
neutron energy. This is often called quasi-free scattering condition. We
choose ten different $q^2$-values as shown in Table~II. The related photon
energy  $\omega$, its three momentum $\mid \vec Q \mid$ 
and the c.m. energy of the final 
three nucleons $E^{c.m.}$ (all evaluated nonrelativistically) are also given. 
For the sake of orientation corresponding relativistic values are included as
well. The comparison of these parameters shows already that at the 
high $q^2$-values relativity cannot
be neglected.

In the following figures, Figs.~\ref{fig5}-\ref{fig7},  
we compare first of all 
results for PWIAS,  ``Full'' and 
calculations with FSI but without MEC`s. On top we add  the  result for
the scattering on a free neutron at rest. This is treated fully relativistically
and will be referred to in the figures as the pure neutron result.
Though we concentrate in this paper on
kinematical regions which are optimal to extract neutron information
we would also like to use the occasion to point to other regions in
phase
space where one can study the reaction mechanism and thus nuclear
dynamics. Therefore we not only show the high energy region of the
knocked out neutron but the observables for all neutron energies, where for
the lower ones the proton contribution in the photon absorption is very
substantial.  This is clearly exhibited by displaying also predictions where 
all electromagnetic proton form factors
are put to zero and thus the photon is absorbed only on the neutron. 

Finally  in PWIAS, which is based on a single nucleon current we
show also results where the nonrelativistic single nucleon
current is replaced by the
fully relativistic one. This idea has been put forward before by
Jeschonnek and Donnelly~\cite{Jeschonnek}. 
Our way to represent  that  relativistic current
 which is ideal for a straightforward extension of the partial wave
representation we use up to now is given in  the Appendix A.

We show the observables 
$A_\parallel = A(\theta^\star= 0^\circ)$ in Fig.~\ref{fig5} and 
$A_\perp = A(\theta^\star= 90^\circ)$ in Fig.~\ref{fig6}.
Furthermore as guidance for experiments we also 
provide the sixfold differential cross section in Fig.~\ref{fig7}.

Lets start with $ A_\parallel$. Roughly spoken the picture is the same for all
$q^2$-values with the exception of the lowest one. The ``Full'' result 
rises quickly from the highest neutron
energy $E_n$ and then with some small oscillations remains essentially
flat towards smaller energies. At the higher end of the neutron energy it
is close to the pure neutron value for $q^2$ = 0.2 ${\rm (GeV/c)}^2$ and higher
momentum transfers. The effect of MEC is most pronounced at the first bump after the
sharp rise. PWIAS is
drastically different in the region of higher neutron energies, except at
the very end, where all curves coincide. Thus FSI should be taken into account
if, because of experimental
reasons, some averaging over neutron energies is needed.
The fully relativistic single nucleon current inserted into a
PWIAS calculation has only  a minor effect at the high neutron energies but it
changes the results at the lower ones above $q^2$= 0.3 ${\rm (GeV/c)}^2$ quite
significantly. For some $q^2$-values we dropped artificially the proton
contribution by switching off all electromagnetic proton form factors.
This leads to a drastic change in PWIAS and the ``Full'' calculation at all
energies (except the very highest ones). The two smallest $q^2$-values are
special, especially $q^2$= 0.05 ${\rm (GeV/c)}^2$, where the ``Full''  
calculation is far away
from the pure neutron result.

In this paper we are mainly concerned with the $G_E^n$ effects in $A_\perp$.
Again we find that 
the rough overall
behavior of the ``Full'' result is similar for all $q^2$-values,
except for the two lowest ones. At the high $q^2$-values
oscillations develop as a function of the neutron 
energy and the effect of
MEC's diminishes. In any case MEC effects are mild and disappear in the
high energy region. But FSI remains important for all $q^2$-values as is
obvious by comparing to the PWIAS results. While the latter ones reach
the pure neutron value at the high energy end the ``Full'' curves stay
always below that value. The effect of the relativistic single nucleon
current is again strongly noticeable at $q^2$= 0.3 ${\rm (GeV/c)}^2$ 
and higher momentum transfers. The
proton contribution is quite significant as shown in some examples. At
the two small $q^2$-values calculations without FSI would  obviously
 be totally meaningless.

Fig.~\ref{fig7} displays the six-fold differential cross section against the
neutron energy for a few examples of $q^2$-values. We see a steep rise
at the high neutron energies due to the $^1S_0$ $t$-matrix pole 
in the pp-subsystem
near zero subsystem energy. Since we did not include the Coulomb force 
the cross section values at the very end might change if that 
approximation can be avoided in the future.
The cross section drops quickly
by orders of magnitudes going to smaller $E_n$-values. At the very low
energies there is again a rise which is due to photon absorption on the
protons, as also shown in the figures. At the very high energy end the
proton contribution is dying out. It is also clear that in all cases
PWIAS is highly insufficient.

Now we focus on the central issue, namely the sensitivity of 
$A_\perp $ and of
$A_\perp / A_\parallel $  with respect to changes 
in $G_E^n$. 
Since the cross section drops rapidly with decreasing neutron 
energy and since only at high energies results of the 3N calculations
can be used to extract neutron information,
we present in Fig.~\ref{fig8} only a high neutron energy range.
We show
results for choosing $G_E^n$ according to the fixed H\"ohler parametrisation
and to the values 1.25 $G_E^n$ and 0.75 $G_E^n$. Since 
$A_\parallel $ is not affected we
display only $A_\perp / A_\parallel $ in Fig.~\ref{fig8}.
There are six curves, three for PWIAS and three for the FULL
calculations. As already noticed in the results for $A_\perp$ we also see 
that FSI can never be neglected. If one regards for instance the range of
about 20 MeV below the highest neutron energy then for $q^2$= 0.35 ${\rm (GeV/c)}^2$
and higher the full dynamics shifts the PWIAS results between 10 and 42 \% .
This is comparable to the signature we are after, namely the
changes of the full result by modifying $G_E^n$ by $\pm $ 25~\%.
At the highest neutron energy these changes start at $\pm $ 32 \% 
for $q^2$= 0.25 ${\rm (GeV/c)}^2$ and decrease slightly to
$\pm $ 27 \% at $q^2$= 0.50 ${\rm (GeV/c)}^2$.
Thus there are even enhancements in the changes of the
ratio $A_\perp / A_\parallel $ against the ones in the variation of $G_E^n$. 
At the lower $q^2$-values PWIAS results would 
be totally meaningless. At $q^2$= 0.20 ${\rm (GeV/c)}^2$
those changes in the ratio increase to $ \pm $ 42 \% and at
$q^2$= 0.15 ${\rm (GeV/c)}^2$ even to $\pm $ 204 \% . 
This drastic increase is of
course caused by the smallness of that specific ratio. At the two
smallest $q^2$-values the sensitivity drops rapidly, $\pm$ 17 \%
at $q^2$= 0.1 ${\rm (GeV/c)}^2$ 
and $\pm $ 2 \% at $q^2$= 0.05 ${\rm (GeV/c)}^2$. The reason
is the strong contribution of the proton as seen in Fig.~\ref{fig6}.
Clearly in all cases the pure neutron value is far off.

One can use the results presented in Fig.~\ref{fig8} to estimate roughly
the error in the $G_E^n$ extraction using only PWIAS. Regarding for instance
the cases $q^2$= 0.3 or 0.35 ${\rm (GeV/c)}^2$ and assuming 
that the experimental
value for $A_\perp / A_\parallel $ measured near the high energy end would lie
on the PWIAS curve (with $G_E^n$ multiplied by the factor 1) then for the ``Full''
calculation to agree with the experimental value one would have to increase 
the $G_E^n$-value by 25 \% and more.
Referred to the pure neutron value this change would be even bigger.
Of course this estimate is very rough since the experimental conditions
leading to averaging have to be taken into account and the magnitude of relativistic
effects are basically unknown but it clearly shows the need of ``Full'' 
calculations for any analysis of such experiments.

\section{Summary and Outlook}
\label{secIV}

We performed Faddeev calculations for the processes 
$\overrightarrow{^3{\rm He}}(\vec{e},e')$ and
$\overrightarrow{^3{\rm He}}(\vec{e},e'n)$
based on the NN force AV18 and consistent MEC's. The asymmetry $A_\perp$
in the inclusive process turned out to be not sensitive enough
to $G_E^n$ to allow its extraction. This is due to the strong
proton contribution. Our studies were performed at 
$q^2$= 0.1 and 0.2 ${\rm (GeV/c)}^2$,
which, however, show a tendency for a decrease of the proton
contribution with increasing four momentum transfer. 
Thus we can not rule out that at higher $q^2$-values 
$A_\perp$ 
might be useful to extract $G_E^n$. Our nonrelativistic approach does not
allow to enter into that realm reliably.

The situation is however favorable  in the neutron knock out process 
$\overrightarrow{^3{\rm He}}(\vec{e},e'n)$
to extract $G_E^n$ information by measuring 
$A_\perp / A_\parallel $. In contrast to
possible expectations FSI corrections are mandatory as documented for
several $q^2$-values up to  the highest  one  which we studied,
$q^2$= 0.5 ${\rm (GeV/c)}^2$. 
Though we entered in  the relativistic domain with
purely nonrelativistic calculations it appears likely that the FSI effects
found are fairly  realistic. Therefore relying on 3N continuum calculations,
whose quality has been tested beforehand in pure 3N scattering
processes~\cite{Report}, one can extract from such measurements 
$G_E^n$ information. There are,
however, still
theoretical uncertainties related to MEC's and of course  relativistic
effects.

As a first step into relativity we used the fully relativistic single
nucleon current operator in a PWIAS calculation and found indeed quite
significant changes, but fortunately not in the high energy end of the
neutron spectrum, which is favorable for the $G_E^n$ extraction.

Improvements in the theoretical framework in the near future are
planned. 3N forces will be included, like it is standard by now in pure
3N scattering (see for instance~\cite{Witala.3NF}) and further
types of MEC's. Of special interest thereby
will be to guarantee consistency to the nuclear forces.

 Besides working
with standard potential models the application of effective field theory
concepts in the form of chiral perturbation theory appears to be very
promising in the low momentum region. This has already started 
in two-, three- and four-nucleon systems including the coupling to the
photon field. For a recent overview and references see~\cite{Epelbaum.prague}.

\acknowledgements

This work was supported by
the Deutsche Forschungsgemeinschaft (J.G.) and by 
the Polish Committee for Scientific Research.
One of us (W.G.) would like to thank the Foundation for Polish
Science for the financial support during his stay in Cracow.
The numerical calculations have been performed
on the Cray T90 of the NIC in J\"ulich, Germany.

\appendix
\section{Implementation of the relativistic single nucleon current}

In this Appendix we show how the relativistic single nucleon current 
is used in our calculations, especially in the context of the 3N system.

The relativistic single nucleon current operator
has the well known form 

\begin{eqnarray}
j^\mu (0) = e \sum\limits_{s s'} \int d {\vec l} \int d {\vec l}' 
\sqrt{{ m \over l_0}}
\sqrt{{ m \over {l_0}'}}
{\bar u}(l' s') \left( F_1 \gamma^\mu + i F_2 \sigma^{\mu \nu} (l' - l)_\nu \right) u(l s) \,
b^\dagger (l' s') b (l s) ,
\label{A1}
\end{eqnarray}
where $ l_0 \equiv \sqrt{ m^2 + {\vec l}^{\ 2}}$, $ {l_0}' \equiv \sqrt{ m^2 + {\vec {l'}}^{\ 2}}$
($m$ is the nucleon mass) and $b^\dagger (l' s')$, $ b (l s) $ are nucleon 
creation and annihilation operators. 
It can be rewritten as 

\begin{eqnarray}
j^\mu (0) = e \sum\limits_{s s'} \int d {\vec l} \int d {\vec l}' 
\sqrt{{ m \over l_0}}
\sqrt{{ m \over {l_0}'}}
{\bar u}(l' s') \left( G_m \gamma^\mu - F_2 (l + l')^\mu \right) u(l s) \,
b^\dagger (l' s') b (l s) \cr
\equiv
e \sum\limits_{s s'} \int d {\vec l} \int d {\vec l}' \,
{\cal X}_{{s}'}^\dagger N^\mu (l, l') {\cal X}_s \, b^\dagger (l' s') b (l s)
\label{A2}
\end{eqnarray}
The last form shows a four-component $2 \times 2$ matrix operator 
acting on Pauli spinors ${\cal X}_s$.
With 
\begin{eqnarray}
A \equiv \sqrt{{ m \over l_0}} \sqrt{{ m \over {l_0}'}} 
\sqrt{{ {{l_0}' + m} \over {2 m} }} \sqrt{{ {{l_0} + m} \over {2 m} }}
\label{A3}
\end{eqnarray}
the components $N^\mu (l, l') $ are written as

\begin{eqnarray}
N^0 = A \left\{ 
\left[ G_m - F_2 (l + l')^0 \right] +
\left[ G_m + F_2 (l + l')^0 \right] { {{\vec l}' \cdot {\vec l}} \over { ({l_0} + m) ({l_0}' + m)} }
\right\} \cr
+ 
A \left[ G_m + F_2 (l + l')^0 \right] 
{ { i {\vec \sigma} \cdot ({\vec l}' \times {\vec l}) } \over { ({l_0} + m) ({l_0}' + m)} }
\label{A4}
\end{eqnarray}

\begin{eqnarray}
N^k = - A F_2 \left(  1 - 
{ {{\vec l}' \cdot {\vec l}} \over { ({l_0} + m) ({l_0}' + m)} }
\right) \,  (l + l')^k \cr
+ A G_m  \left( { l^k \over {{l_0} + m} } + { {l'}^k \over {{l_0}' + m} } \right) \cr
+ A F_2 { { (l + l')^k  } \over { ({l_0} + m) ({l_0}' + m)} } \, 
i {\vec \sigma} \cdot ({\vec l}' \times {\vec l}) \cr
+ A G_m \left[ { 1 \over {({l_0} + m)} } \, i ( {\vec l} \times {\vec \sigma} )^k 
             + { 1 \over {({l_0}' + m)} } \, i ( {\vec \sigma} \times {\vec {l'}} )^k
        \right]
\label{A5}
\end{eqnarray}

Introducing standard Jacobi momenta $ {\vec p} , {\vec q} $
the current matrix element between initial
$ \varphi $ and final $ {\varphi}' $ 3N states can be written as

\begin{eqnarray}
\langle {\varphi}' \{ M' \} {\vec {{\cal P}'}} \mid  j^\mu (0) \mid 
        {\varphi}     M     {\vec {\cal P} } \rangle \cr
= \, 3 
\int d {\vec p} \int d {\vec q} \,
\langle {\varphi}' \{ M' \} \mid {\vec p} ,  {\vec q} \rangle \, 
N^\mu (l' , l) \,
\langle {\vec p} , {\vec q} + \frac23 {\vec {\cal P} } - \frac23 {\vec {{\cal P}'}} 
\mid {\varphi} M  \rangle  , 
\label{A6}
\end{eqnarray}
where $ {\vec {l'}} \equiv {\vec q} + \frac13 {\vec {{\cal P}'}} , 
        {\vec {l }} \equiv {\vec q} + {\vec {\cal P} } - \frac23 {\vec {{\cal P}'}}$.
$ {\vec {\cal P} }$ and ${\vec {{\cal P}'}}$ are the initial and final 
total 3N momenta,
respectively. 

We choose the laboratory frame (${\vec {\cal P} }= 0 $) and denote 
$ {\vec Q} =  {\vec {{\cal P}'}} - {\vec {{\cal P} }} = {\vec {{\cal P}'}} $.
Furthermore because of current conservation 
we can restrict ourselves only to transverse components of $N^k$,
and choose the spherical components 
$N_\tau, \tau = \pm 1$.
Then expressions appearing in Eqs.~(\ref{A4}) and (\ref{A5}) can be evaluated as

\begin{eqnarray}
{\vec l}' \cdot {\vec l} = q^2 - \frac13 {\vec q} \cdot {\vec Q} - \frac29 Q^2 \cr
{\vec l}' \times {\vec l} = {\vec Q} \times {\vec q} \cr
l_\tau = {l'}_\tau = q_\tau  \cr
( {\vec \sigma} \times {\vec {l'}} )_\tau = 
( {\vec \sigma} \times {\vec {q}} )_\tau  + \frac13 
( {\vec \sigma} \times {\vec {Q}} )_\tau  \cr
( {\vec \sigma} \times {\vec {l}} )_\tau = 
- ( {\vec \sigma} \times {\vec {q}} )_\tau  + \frac23 
( {\vec \sigma} \times {\vec {Q}} )_\tau 
\label{A7}
\end{eqnarray}
and one can group some terms in Eq.~(\ref{A5}) together. One ends up with

\begin{eqnarray}
N_\tau =  
A \left\{  
G_m  \left( { 1 \over {{l_0} + m} } + { 1 \over {{l_0}' + m} } \right)
- 2 F_2 \left(  1 - { {{\vec l}' \cdot {\vec l}} \over { ({l_0} + m) ({l_0}' + m)} } \right) 
\right\} \, q_\tau  \cr
+ A G_m \left( 
{ {\frac23} \over {({l_0} + m)} } + { {\frac13} \over {({l_0}' + m)} } 
       \right) \, i ( {\vec \sigma} \times {\vec Q} )_\tau \cr
+ A G_m \left( 
{ {1} \over {({l_0} + m)} } - { {1} \over {({l_0}' + m)} } 
       \right) \, i ( {\vec q} \times {\vec \sigma} )_\tau \cr
+ A 2 F_2 { { q_\tau  } \over { ({l_0} + m) ({l_0}' + m)} } \,
i {\vec \sigma} \cdot ({\vec Q} \times {\vec q}) 
\label{A8}
\end{eqnarray}

In the nonrelativistic limit only the correspondingly reduced first two terms 
in Eq.~(\ref{A8}) remain;
the first one is the convection current, the second is the spin current.
The partial wave decomposition can be carried through by
straightforward extension of the forms given in~\cite{Golak.95}. 
As a subtle point
we mention that the arguments of the electromagnetic form factors are
not the four-momentum squared of the photon but 
$ ({l_0} - {l_0}')^2 - ({\vec l} - {\vec l}')^2 $.
This is required in a Hamiltonian formalism where only the three-momenta
are conserved at the vertices and not the four-momenta as in a manifest
covariant formalism.


\begin{table}[hbt]
\begin{center}
\begin{tabular}{lcccc} \hline

             &  \multicolumn{4}{c}{\rule[-3mm]{0mm}{10mm}
                  $q^2$= 0.1 ${\rm (GeV/c)}^2$, $\omega$= 50 MeV}   \\
             &         $R_L$        &        $R_T$           &       $R_{T'}$        & $R_{TL'}$   \\ \hline
``Full''         & 1.91$\times$10$^{-2}$ & 1.07$\times$10$^{-2}$ & 1.86$\times$10$^{-3}$ & 9.68$\times$10$^{-4}$ \\
``Full'' (no proton)  & 1.19$\times$10$^{-5}$ & 1.38$\times$10$^{-3}$ & 1.35$\times$10$^{-3}$ & 1.24$\times$10$^{-4}$ \\ 
\hline 

             &  \multicolumn{4}{c}{\rule[-3mm]{0mm}{10mm}
                  $q^2$= 0.2 ${\rm (GeV/c)}^2$, $\omega$= 110 MeV}   \\
             &         $R_L$        &        $R_T$           &       $R_{T'}$        & $R_{TL'}$   \\ \hline
``Full''         & 1.04$\times$10$^{-2}$ & 1.06$\times$10$^{-2}$ & 1.72$\times$10$^{-3}$ & 7.82$\times$10$^{-4}$ \\
``Full'' (no proton)  & 1.81$\times$10$^{-5}$ & 1.47$\times$10$^{-3}$ & 1.42$\times$10$^{-3}$ & 1.57$\times$10$^{-4}$ \\ 
\end{tabular}
\end{center}
\caption[]
{
Response functions for inclusive scattering and for two $q^2$-values 
at $\omega$-values in the peak region. The ``Full'' calculation is compared
to calculations without absorption of the photon on the proton.
All responses are given in the units of 1/MeV.
}
\label{TAB2}
\end{table}

\begin{table}[htb]
\label{TAB1}
\begin{tabular}{ccccccc}
\hline
$q^2$ & $\omega_{\rm nrl}$ & $Q_{\rm nrl}$ & $\omega_{\rm rel}$ & $Q_{\rm rel}$ & $E^{c.m.}_{\rm nrl}$ &
$E^{c.m.}_{\rm rel}$ \\
${\rm (GeV/c)}^2$ & MeV & MeV/c & MeV & MeV/c & MeV & MeV \\
\hline
         0.05  &    27.0 & 225.2 &    26.6 & 225.2 &    12.5 &   12.2  \\
         0.10  &    54.8 & 320.9 &    53.2 & 320.7 &    31.1 &   29.7  \\
         0.15  &    83.6 & 396.2 &    79.9 & 395.4 &    50.3 &   47.2  \\
         0.20  &   113.3 & 461.6 &   106.5 & 459.7 &    70.1 &   64.5  \\
         0.25  &   144.2 & 520.4 &   133.1 & 517.4 &    90.6 &   81.8  \\
         0.30  &   176.3 & 575.4 &   159.7 & 570.5 &   112.0 &   98.9  \\
         0.35  &   209.8 & 627.7 &   186.4 & 620.3 &   134.4 &  116.0  \\
         0.40  &   244.9 & 678.2 &   213.0 & 667.4 &   157.8 &  132.9  \\
         0.45  &   281.9 & 727.7 &   239.6 & 712.3 &   182.5 &  149.7  \\
         0.50  &   321.1 & 776.6 &   266.2 & 755.6 &   208.6 &  166.5  \\
\hline
\end{tabular}
\caption[]{Kinematical quantities for quasi-free scattering conditions
           studied in the present work. The electron beam energy was fixed to 1 GeV.}
           Subscripts ``nrl'' and ``rel'' refer to the nonrelativistic and relativistic 
           treatment of kinematics.
\end{table}


\begin{figure}[h!]
\leftline{\mbox{\epsfysize=185mm \epsffile{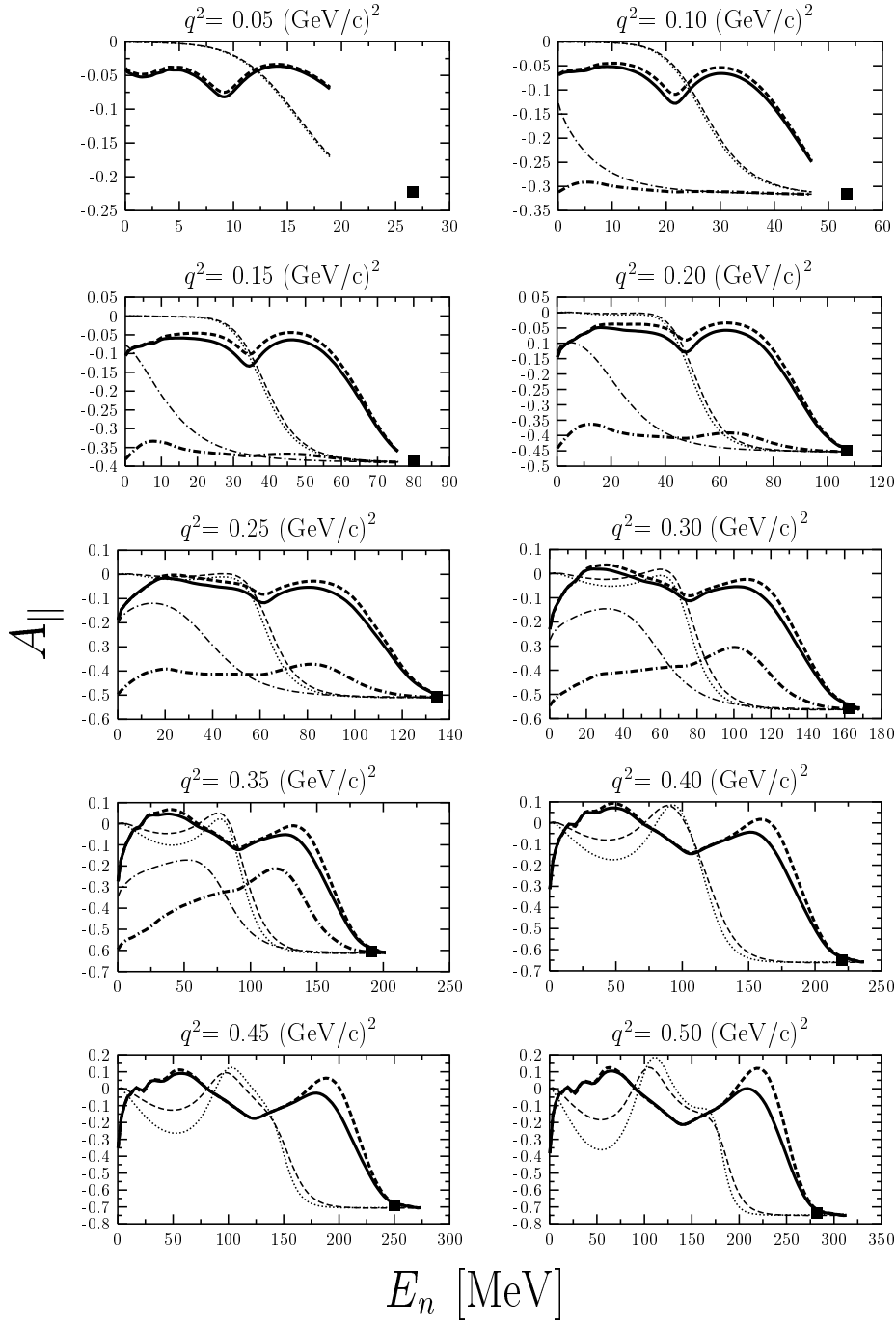}}}
\caption[ ]
{
$A_\parallel$ as a function of the neutron energy $E_n$
for different $q^2$-values.
``Full'' (solid), FSI without MEC (dashed, thick line), 
``Full'' without proton contribution (dash-dotted, thick line), 
PWIAS (dashed, thin line), PWIAS with the relativistic single nucleon 
current (dotted) and PWIAS without proton contribution (dash-dotted, thin line);
pure neutron result (filled square).
The dashed-dotted lines occur only for $q^2$ from 0.1 to 0.35 ${\rm (GeV/c)}^2$.
}
\label{fig5}
\end{figure}

\begin{figure}[h!]
\leftline{\mbox{\epsfysize=185mm \epsffile{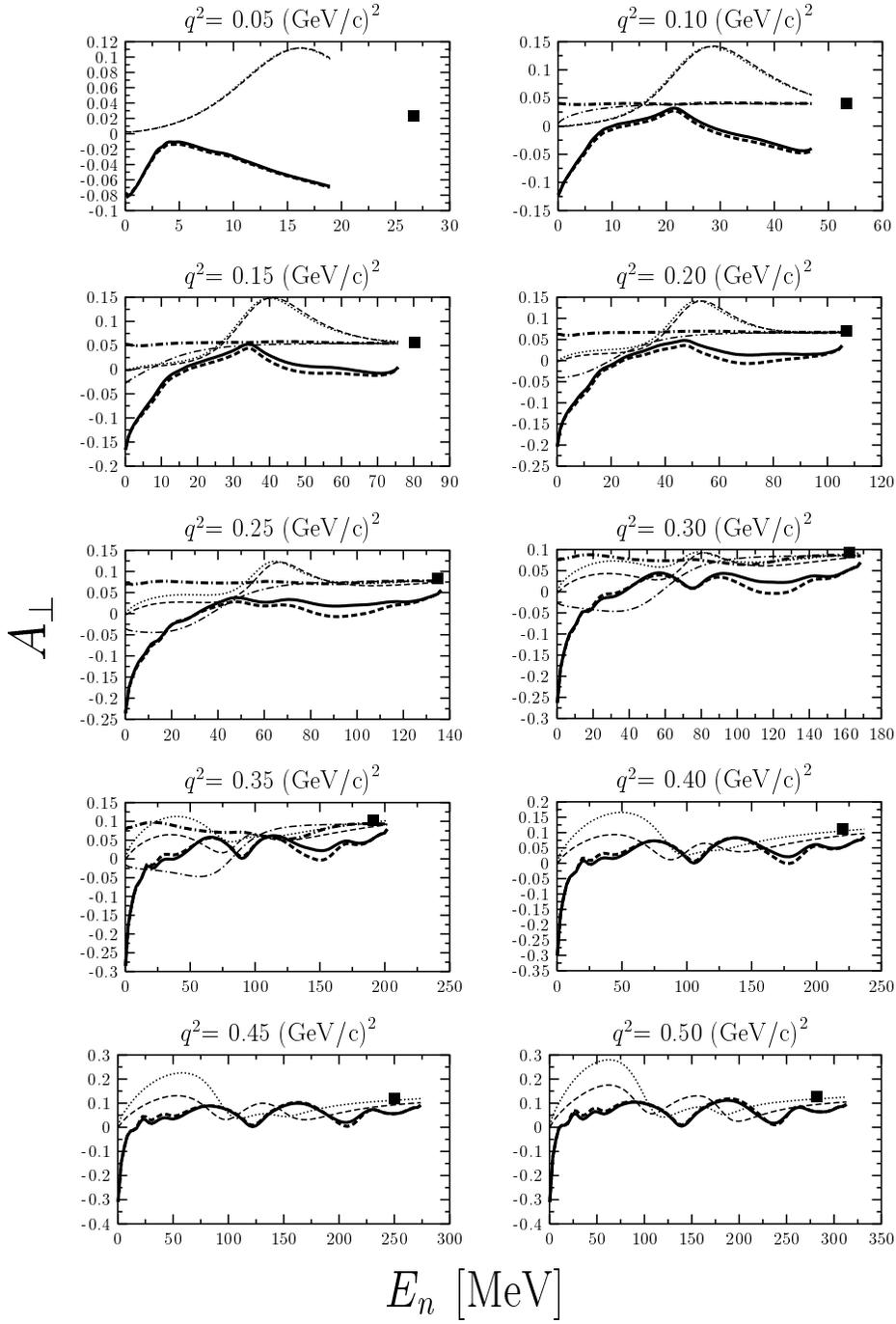}}}
\caption[ ]
{
$A_\perp$ as a function of the neutron energy $E_n$
for different $q^2$-values. Curves and the symbol 
as in Fig.~\protect\ref{fig5}.
}
\label{fig6}
\end{figure}

\begin{figure}[h!]
\leftline{\mbox{\epsfxsize=120mm \epsffile{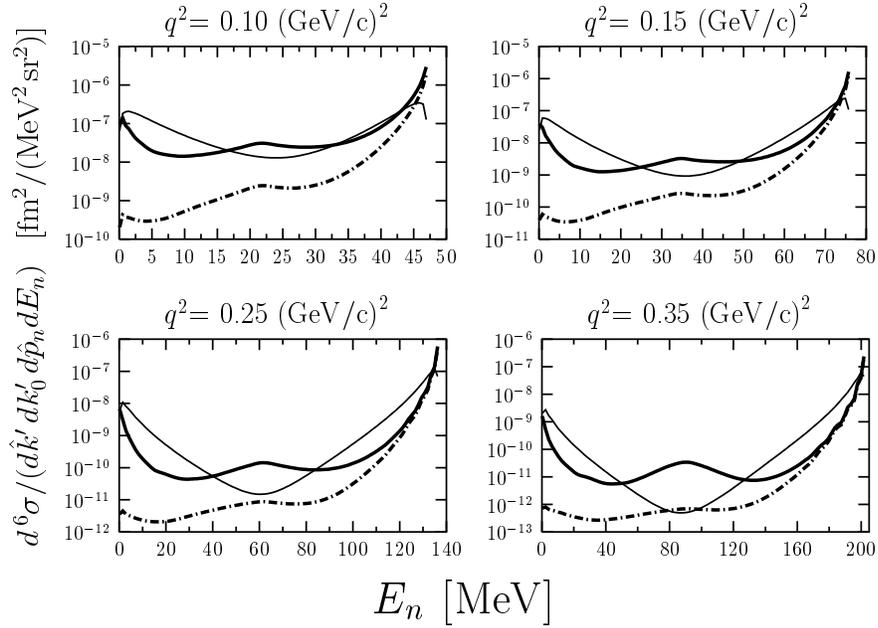}}}
\caption[ ]
{
The sixfold differential cross section
as a function of the neutron energy $E_n$
for different $q^2$-values. 
``Full'' (solid, thick line), PWIAS (solid, thin line) and ``Full'' without 
proton contribution (dash-dotted).
}
\label{fig7}
\end{figure}

\begin{figure}[h!]
\leftline{\mbox{\epsfysize=185mm \epsffile{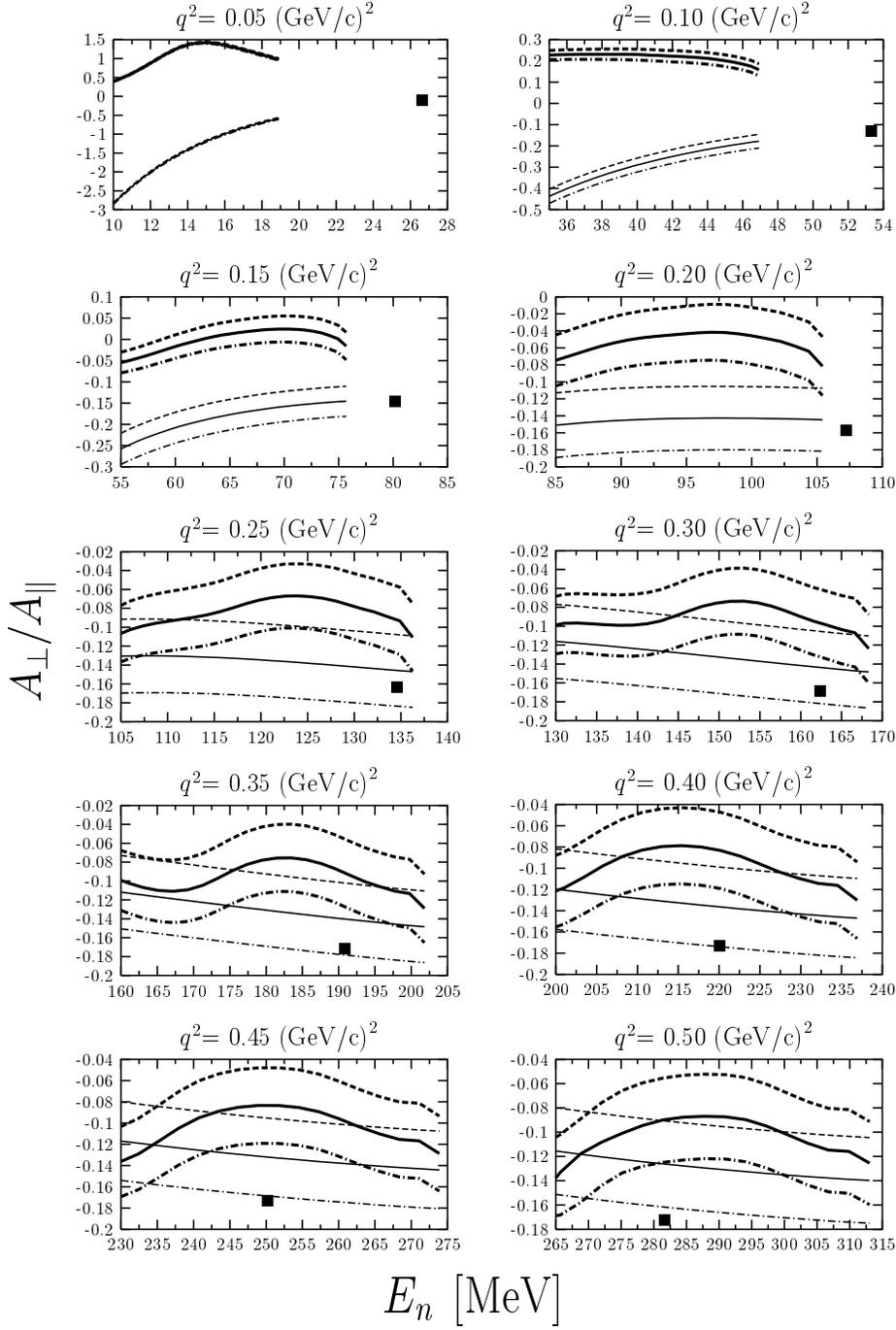}}}
\caption[ ]
{
The ratio $A_\perp / A_\parallel $
as a function of the neutron energy $E_n$
for different $q^2$-values. The thick lines are: 
``Full'' with 1.0$G_E^n$ (solid), ``Full'' with 0.75$G_E^n$ (dashed)
and ``Full'' with 1.25$G_E^n$ (dash-dotted). The thin lines are the corresponding 
cases for PWIAS.
Filled square is the pure neutron result.
}
\label{fig8}
\end{figure}

\end{document}